\documentclass[conference, 10pt]{IEEEtran}

\usepackage{cite}
\usepackage{url}
\usepackage{comment}
\usepackage{graphicx}
\usepackage{color}
\usepackage{placeins}
\usepackage{float}
\usepackage{tabularx,colortbl}
\usepackage{ifthen}
\usepackage{tikz}
\usepackage{tikz-3dplot}
\usepackage{subcaption}
\usepackage{circuitikz}
\usetikzlibrary{shapes.geometric}
\usepackage{amsmath,amsfonts,amsthm,amssymb,esint} 
\usepackage{subcaption}
\usepackage[
top    = 21mm,
bottom = 1.3in,
left   = 0.62in,
right  = 0.62in]{geometry}
\definecolor{darkblue}{rgb}{0.0,0.0,0.3}
\definecolor{darkred}{rgb}{0.3,0.0,0.0}
\definecolor{darkgreen}{rgb}{0.0,0.3,0.0}

\makeatletter

\newcounter{author}
\renewcommand{\author}[2][]{
   \stepcounter{author}
   \@namedef{author@\theauthor}{#2}
   \@namedef{authorlabel@\theauthor}{#1}
}

\newcounter{address}
\newcommand{\address}[2][]{
   \stepcounter{address}
   \@namedef{address@\theaddress}{#2}
   \@namedef{addresslabel@\theaddress}{#1}
}

\newcommand{\alsep}{and}

\def\newmaketitle{\par%
  \begingroup%
  \normalfont%
  \def\thefootnote{}
  \def\footnotemark{}
  \let\@makefnmark\relax
  \footnotesize
  \footnotesep 0.7\baselineskip
  \normalsize%
  \twocolumn[\thenewmaketitle\@IEEEaftertitletext]%
  \if@IEEEusingpubid
     \enlargethispage{-\@IEEEpubidpullup}%
  \fi
  \endgroup
  \setcounter{footnote}{0}\let\maketitle\relax\let\@maketitle\relax
  \gdef\@thanks{}%
  \let\thanks\relax}

\def\thenewmaketitle{
  \newpage
  \begin{center}%
    \vskip0.2em{\Huge\@IEEEcompsoconly{\sffamily}\@IEEEcompsocconfonly{\normalfont\normalsize\vskip 2\@IEEEnormalsizeunitybaselineskip
   \bfseries\large}\@title\par}\vskip1.0em\par%
    \vspace{1ex}
    \newcounter{c@author}
    \newcounter{c@tmp}
    \ifthenelse{\value{author}=2}{%
      \newcommand{\liand}{ and }}{%
      \newcommand{\liand}{, and }}
    \ifthenelse{\value{address}<2}{%
      \@nameuse{author@1}%
      \stepcounter{c@author}%
      \whiledo{\value{c@author}<\value{author}}{%
        \setcounter{c@tmp}{\value{author}}%
        \addtocounter{c@tmp}{-\value{c@author}}%
        \ifthenelse{\value{c@tmp}=1}{%
          \renewcommand{\alsep}{\liand}}{\renewcommand{\alsep}{, }}%
        \stepcounter{c@author}\alsep \@nameuse{author@\thec@author}}\\%
    }
    {
      \@nameuse{author@1}${}^{(\ref{\@nameuse{authorlabel@1}})}$%
      \stepcounter{c@author}%
      \whiledo{\value{c@author}<\value{author}}{%
      \setcounter{c@tmp}{\value{author}}%
      \addtocounter{c@tmp}{-\value{c@author}}%
      \ifthenelse{\value{c@tmp}=1}{%
        \renewcommand{\alsep}{\liand}}{\renewcommand{\alsep}{, }}%
      \stepcounter{c@author}\alsep \@nameuse{author@\thec@author}%
        ${}^{(\ref{\@nameuse{authorlabel@\thec@author}})}$%
      }
    }
    \vspace{0.2ex}

    \ifthenelse{\value{address}>0}{%
      \ifthenelse{\value{address}=1}{
        {\@nameuse{address@1}}
      }
      {
        \newcounter{c@address}

        \begin{center}
        \whiledo{\value{c@address}<\value{address}}
        {
          \refstepcounter{c@address}
            ${}^{(\thec@address)}$\,%
              \label{\@nameuse{addresslabel@\thec@address}}%
              \@nameuse{address@\thec@address}\\ %
        }
        \end{center}
      } 
    }
    {
      \relax
    }
  \end{center}
}

\makeatother

\title{Physical Origin of the Effective Parameters at Boundaries in Finite Difference Schemes}

\author[org1]{Sameh Y.~Elnaggar}
\author[org1]{Yahia M.~M.~Antar}

\address[org1]{The Royal Military College of Canada,\\ Email: S.Y.E (samehelnaggar@ieee.org), Y.~M.~M.~A (antar-y@rmc.ca)}

\begin{document}

\newmaketitle

\begin{abstract}
Using the Observable form of Maxwell's equations, we reveal that effective parameters at materials boundaries emerge naturally as  anisotropic  transfer functions. The complexity of the boundary dictates the order of these functions. Employing the residue-pole expansion, we describe the relation between field components as a system of ODEs solvable through the auxiliary differential equation approach. Not only can be used in Finite difference methods, the approach can be also employed to model complex surface structures.
\end{abstract}

\section{Introduction}
Finite Difference (FD) approaches stand out as straightforward and intuitive methods for solving Maxwell's equations and provide an excellent opportunity to introduce numerical methods to students. In the standard FD approach, the computational domain is discretized into staggered cells, known as Yee cells \cite{ Taflove2000}. However,  challenges arise when dealing with material objects that do not align perfectly with the cell faces.

Sampling material parameters at cell centers, creating a staircase representation, is a common approach to model materials that, however, compromises solution accuracy. Various techniques have been proposed to address these limitations and better model material boundaries \cite{nadobny2008general, zhao2019generalized, oskooi2009accurate}. (For a comprehensive review, refer to Ref. \cite{glytsis2018review}.) However, many of these approaches appear (at first sight) mathematically complex,  lack a direct connection to the underlying physics and hence make the introduction of the FD methods challenging to new students.

In this article, we demonstrate that rigorously defining effective parameters at boundaries is achievable by adopting the observable formalism of Maxwell's equations. Besides being easy to grasp, the observable formalism, as embraced by the authors \cite{Sameh_EMTS2023} and discussed by other researchers \cite{clemens2001discrete, tonti2013mathematical}, offers an intuitive interpretation of finite difference methods.

Furthermore, we present a straightforward procedure for describing effective parameters in the time domain using auxiliary differential equations.

\section{Theory}
\subsection{Observable Formalism and Obsverable Quantities}
Material inclusion involves averaging over the microscopic scale to calculate macroscopic fields. This process spans dimensions large enough to encompass hundreds or thousands of atoms but remains infinitesimal from an observer's standpoint.

On the other hand, measurements are conducted using equipment that senses average quantities over a larger scale. A primary purpose of a computational tool is to predict measurement results. In this context, the finiteness of a computational tool is regarded as a representation of the observable scale, rather than the underlying set of differential equations \cite{tonti2013mathematical}.

Maxwell's equations lend themselves to an observable form, which reduces to the well-known differential form in the limit of infinitesimal dimensions. For example, the two circuital laws can be formulated in an observable form via the integration of the electromotive force (EMF) and magnetomotive force (MMF) over some time intervals \([t_1, t_2]\):\vspace{-3mm}

\begin{align*}
\sum_\circlearrowleft\mathcal{E}=-\left(\Psi(t_2)-\Psi(t_1)\right),~~
\sum_\circlearrowleft\mathcal{M}=\left(\Phi(t_2)-\Phi(t_1)\right)+Q_f,
\end{align*}
where \(\sum_\circlearrowleft\mathcal{E}\) (\(\sum_\circlearrowleft\mathcal{M}\)) represents the integration of the EMF (MMF) from $t_1$ to $t_2$. \(\Psi(t)\) (\(\Phi(t)\)) is the total magnetic (electric) flux crossing the surface at time \(t\), and \(Q_f\) is the total charge flowing through the surface over the time interval.

It is noteworthy that the observable form allows for an intuitive interpretation of finite difference approaches and is linked to the physics \cite{Sameh_EMTS2023, clemens2001discrete, tonti2013mathematical}.

\begin{figure}
\tdplotsetmaincoords{70}{145}
\begin{subfigure}{0.4\linewidth}
\begin{tikzpicture}[tdplot_main_coords,scale=3]
\draw [gray,thick,->](0,0,0)--(0,0,1) node[anchor=north west]{$z$} ;
\draw [gray,thick,->] (0,0,0)--(1,0,0) node[anchor=east]{$x$};
\draw [gray,thick,->] (0,0,0)--(0,1,0) node[anchor=west]{$y$};
\draw [ultra thick,darkblue,fill=darkblue,fill opacity=0.3](0.5,0.5,0.5)--(0.5,0.0,0.5) node[
    currarrow,
    pos=0.5, 
    xscale=-1,
    sloped,
    scale=1] {}--(0.5,0,0.2)node[
    currarrow,
    pos=0.5, 
    xscale=1,
    sloped,
    scale=1] {}--(0.5,0.5,0.2) node[
    currarrow,
    pos=0.5, 
    xscale=1,
    sloped,
    scale=1] {}--(0.5,0.5,0.5) node[
    currarrow,
    pos=0.5, 
    xscale=1,
    sloped,
    scale=2] {};
    
    \draw [darkred,ultra thick,dashed] ((0.2,0.25,0.35)--(0.5,0.25,0.35) ;
    
  \draw [darkred, ultra thick, fill opacity=0.3](0.5,0.25,0.35)--(0.75,0.25,0.35) node[
    currarrow,
    pos=0.5, 
    xscale=-1,
    sloped,
    scale=1] {} --(0.75,0.25,0.65) node[
    currarrow,
    pos=0.5, 
    xscale=1,
    sloped,
    scale=1] {} --(0.2,0.25,0.65) node[
    currarrow,
    pos=0.5, 
    xscale=1,
    sloped,
    scale=1] {} --(0.2,0.25,0.35)node[
    currarrow,
    pos=0.5, 
    xscale=1,
    sloped,
    scale=1] {}  ;
    
    \draw [red,fill=red] node at (0.5,0.25,0.35){\LARGE .};
\end{tikzpicture}
\caption{}
\end{subfigure}
\begin{subfigure}{0.4\linewidth}
\centering
\tdplotsetmaincoords{70}{155}

\begin{tikzpicture}[tdplot_main_coords,scale=1.4]
\draw[gray,->](0,0,0)--(1.5,0,0)node[anchor=east]{$x$};
\draw[gray,->](0,0,0)--(0,1.5,0)node[anchor=west]{$y$};
\draw[gray,->](0,0,0)--(0,0,1.5)node[anchor=west]{$z$};
\draw [ultra thick,blue, fill=blue, opacity=0.15](0,0,0)--(1,1,0)--(1,1,1)--(0,0,1)--cycle;
\draw [ultra thick,blue](0,0,0)--(1,1,0)--(1,1,1)--(0,0,1)--cycle;
\draw [ultra thick, red](0.5,0.5,0.5)--(0,1,0.5);
\draw [ultra thick, red, dashed](0.5,0.5,0.5)--+(0.5,-0.5,0);
\draw [ultra thick,->, red](0.5,0.5,0.5)--+(-0.1,0.1,0);
\draw [blue](0.2,0,0.6) node{$\Phi$};
\draw [red](0,1.2,0.6)node{$V$};
\end{tikzpicture}
\caption{}
\end{subfigure}
\caption{(a) Two staggered surfaces. $E_x$ is normal to the blue surface, which is at the same time tangential to one of the edges of the red surface. (b) Capacitance at face=$\Phi/V$.\vspace{-2mm}}
\label{fig:staggered}
\end{figure}
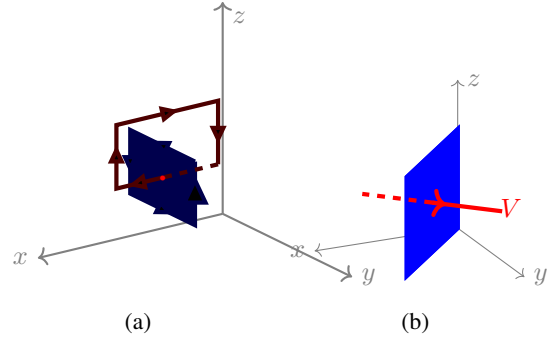

\subsection{Observable effective parameters in Frequency Domain}
To establish a connection between the two circuital equations, a staggered configuration, as illustrated in Fig. \ref{fig:staggered}, is implicitly assumed. It naturally leads to relating the electric flux (or its areal density $\langle D_x\rangle_s$) over the filled surface to the electromotive force (EMF) (or its linear density $\langle E_x\rangle_l$) along the line intersecting it. This relationship is a constitutive relation (in fact $\langle D_x\rangle_s/\langle E_x\rangle_l$ is the capacitance per unit length). In the infinitesimal limit, it reduces to $D_x=\epsilon E_x$. In the general case, the relation becomes more intricate:\vspace{-1mm}
\begin{equation}
\langle \mathbf{D}\rangle_s=\bar{\mathbf{\epsilon}}(\omega)\langle \mathbf{E}\rangle_l,
\end{equation}
where $\mathbf{\epsilon}$ is an observable dielectric tensor-like operator, $\langle \mathbf{D}\rangle_s=\left(\langle D_x\rangle_s, \langle D_y\rangle_s, \langle D_z\rangle_s\right)$, and $\langle \mathbf{E}\rangle_l=\left(\langle E_x\rangle_l, \langle E_y\rangle_l, \langle E_z\rangle_l\right)$. The tensorial relation directly stems from the continuity of the normal (tangential) component of the \emph{macroscopic} $\mathbf{D}$ ($\mathbf{E}$).

\begin{figure}
\centering
\begin{tikzpicture}[scale=1]
\draw [ultra thick, blue](-1,0)--(1,0) node[anchor=west]{Yee face};
\draw [ultra thick, dashed, red](0,-0.75)--(0,0.75);
\draw [gray, thick](0.5,-0.2)--(1.5,1.5) (0.5,-0.2)--(-1,-1);
\draw [teal, thick,->](-0.5,0)--(-0.5,-0.3)node[anchor=east]{ $\mathbf{\hat{s}}$};
\draw [teal, ->](-0.6,-0.8)--+(1.2*0.125,-1.2*0.2)node[anchor=north west]{ $\mathbf{n_y}$};
\draw [teal, ->](-0.6,-0.8)--+(1.6*0.125,-0)node[anchor=north west]{ $\mathbf{n_{yx}}$};
\draw [teal, ->](-0.6,-0.8)--+(0,-1.4*0.2)node[anchor=north east]{$\mathbf{n_{yy}}$};
\draw [teal,  ->](0.75,3.5*0.1)--+(0.2*1.7,-0.2*1);
\draw [teal, ->](1.2*0.67,3.5*0.1)--+(0.2*1.7,0)node[anchor=south west]{$\mathbf{n_{xx}}$};
\draw [teal, ->](0.75,0.35)--+(0,-0.25*1);
\node [teal] at(0.75,0.35)[anchor= east]{$\mathbf{n_{xy}}$};

\node [anchor=south east]{ $O$};
\node at(0.2,,0.06){$\epsilon_O$};
\end{tikzpicture}
\caption{Two interfaces case, with different parameters shown.}
\label{fig:two_interfaces}
\end{figure}
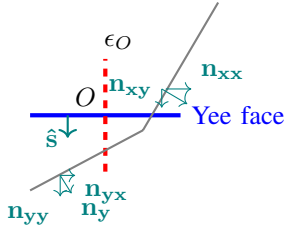
To demonstrate how $\bar{\epsilon}$ may look like, we consider the 2D case shown in Fig. \ref{fig:two_interfaces}. The material surface intersecting with the Yee face centered at point $O$ has a normal vector $(n_{xx},n_{xy})$. We also consider the intersection with the normal segment (dashed line in the figure) to be generally different $(n_{yx},n_{yy})$. Utilizing the continuity of normal $\mathbf{D}$ and tangential $\mathbf{E}$ across interfaces and assuming that $\epsilon$ is complex (including conductivity), it can be shown that $\langle \mathbf{D}\rangle_s=\left[\mathbf{\mathcal{H}}\right]\langle \mathbf{E}\rangle_l$,
where $\mathcal{H}_{ij}(s)$ is a transfer function that can be written in residue-pole form:\vspace{-1mm}
\begin{equation}
\mathcal{H}_{ij}(s)=\epsilon_{ij}+\frac{\sigma_{ij}}{s}+\frac{A_{1_{ij}}}{s-p_1}+\frac{A_{2_{ij}}}{s-p_2},
\end{equation}
where $\epsilon_{ij}$ and $\sigma_{ij}$ are the effective observable dielectric constant and conductivity, respectively. The second two terms appear as $e^{-p_it}u(t)$ in the time domain, emphasizing the fact that the change in $\langle E\rangle_l$ results in a delayed response in $\langle D\rangle_s$ due to the charge accumulating at the object surfaces (analog to an RC circuit). Note that when $n_{xx}=n_{yx}, n_{xy}=n_{yy}$, then $A_{2_{ij}}=0$. Each intersection accounts for a residue-pole term and is analog to an RC circuit. Magnetic parameters can be determined in a similar fashion.
\subsection{Auxiliary Differential Equations}
In the time domain, the relation between the fields is obtained via the application of inverse Fourier transform. For simplicity of notation we denote $\langle D_x\rangle_s$ by $D_x$ and $\langle E_w\rangle_l$ by $E_w$. Therefore,\vspace{-2mm}
\begin{align*}
\partial_tD_x&=\sum_{w=x,y}\epsilon_{xw}\partial_tE_w+\sigma_{xw}E_w+\mathcal{J}_{1w}(t)+\mathcal{J}_{2w}(t).
\end{align*}
Each current-like term $\mathcal{J}_{i\alpha}$ can be related to the fields via a first order differential equation. For example:\vspace{-2mm}
\begin{equation}
\label{eq:relax_current}
\mathcal{J}_{1x}-p_1\partial_t\mathcal{J}_{1x}=A_{1_{xx}}\partial_tE_x.
\end{equation}
Equation (\ref{eq:relax_current}) can be discretized at the time instant $q\Delta_t$, where $\mathcal{J}_{1x}$ can be represented in terms of $E_x$.

\section{Results}
\begin{figure}
\includegraphics[scale=0.3]{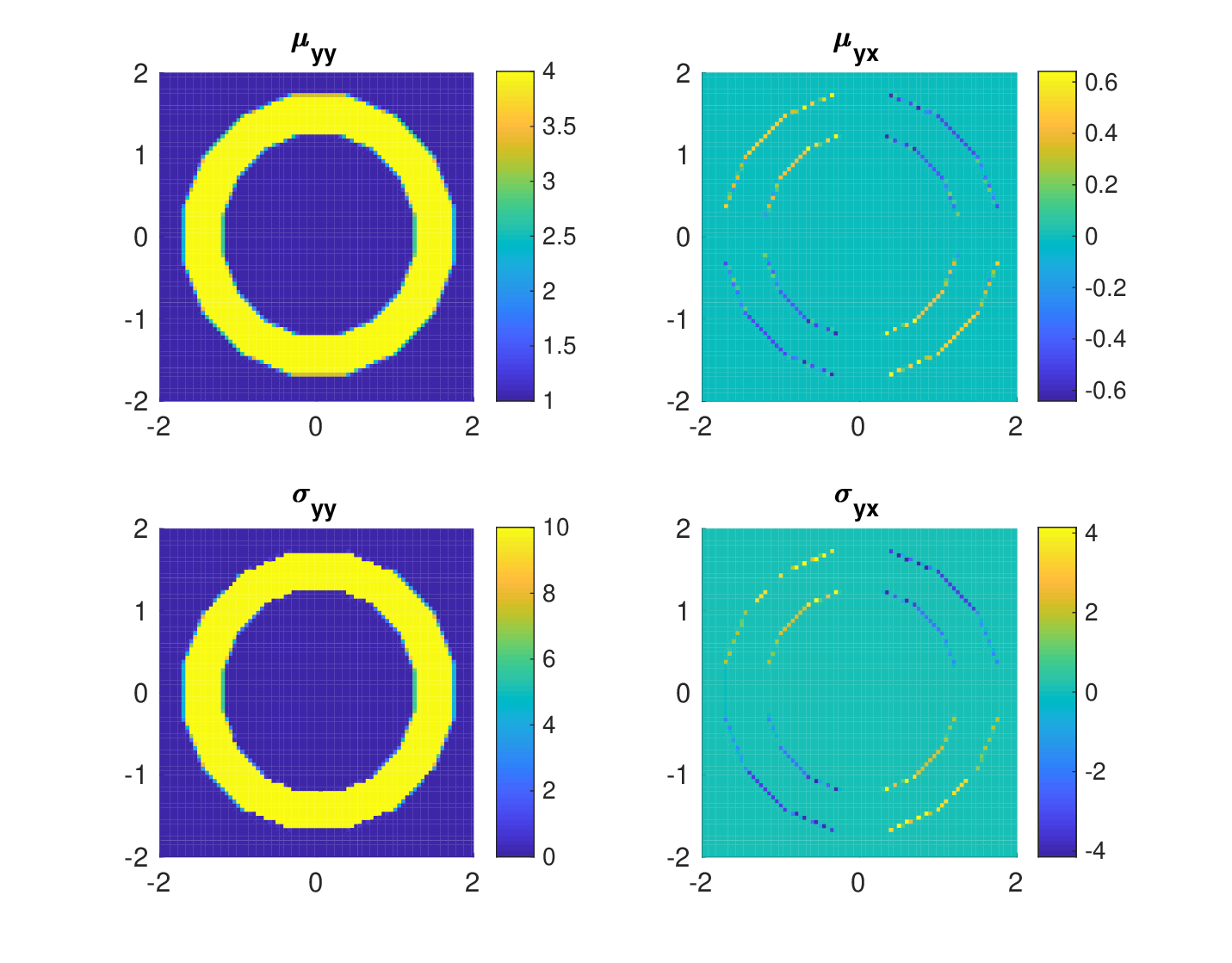}
\caption{Effective Parameters of circular ring, with bulk parameters: $\epsilon_r=4, \sigma=10, \mu_r=4$ and magnetic conductivity $\sigma^m=10$, where the normalized units were used ($\epsilon_0=\mu_0=1$).\vspace{-3mm}}
\label{fig:results}
\end{figure}

To ilustrate the above procedure, a 2D cylindrical ring with the effective parameters presented in Fig. \ref{fig:results} is employed. For a TM polarization ($H_z=0$), the pertinent parameters include $\epsilon_{zz}$, $\sigma_{zz}$, and the in-plane magnetic permeability and conductivity ($\mu_{xx}$, $\mu_{xy}$, $\mu_{yx}$, and $\mu_{yy}$). These parameters are visually represented in Fig. \ref{fig:results}, where it is notable that the off-diagonal terms can be negative based on the sign of $n_xn_y$.
Furthermore, ongoing work involves a comparison with the staircase approach, and the results of this comparison will be presented in the conference.

Our discussion highlights that anisotropy and frequency dependency at the boundaries naturally emerge from the Observable formal of Maxwell's equations.
\bibliographystyle{IEEEtran}
\bibliography{CEM}
\end{document}